\documentclass[aps,prx,amsmath,amssymb,lengthcheck,showpacs,superscriptaddress,floatfix]{revtex4-2}
\usepackage[normalem]{ulem}
\usepackage{graphicx}
\usepackage{color}
\usepackage{datetime}

\newcommand{\rr}{{\bf r}}

\newcommand{\VV}{{\mathcal V}}

\newcommand{\be}{\begin{equation}}
\newcommand{\ee}{\end{equation}}
\newcommand{\ba}{\begin{eqnarray}}
\newcommand{\ea}{\end{eqnarray}}
\newcommand{\bse}{\begin{subequations}}
\newcommand{\ese}{\end{subequations}}
\newcommand{\beq}{\begin{eqnarray}}
\newcommand{\eeq}{\end{eqnarray}}

\newcommand{\ts}{\textstyle}

\begin{document}
\title{Classical versus quantum Anderson localization in disordered systems}
\author{Stefano Mossa}
\affiliation{Univ. Grenoble Alpes, CEA, IRIG-MEM-LSim, 38054 Grenoble, France}
\author{Giancarlo Ruocco}
\affiliation{Center for Life Nano Science @Sapienza, Istituto Italiano di Tecnologia, 295 Viale Regina Elena, I-00161, Roma, Italy}
\affiliation{Dipartimento di Fisica, Universit\'a di Roma "La Sapienza", P.le Aldo Moro 5, I-00185, Roma, Italy}
\author{Walter Schirmacher}
\affiliation{Institut f\"ur Physik, Universit\"at Mainz, Staudinger Weg 7, D-55128 Mainz, Germany}
\affiliation{Center for Life Nano Science @Sapienza, Istituto Italiano di Tecnologia, 291 Viale Regina Elena, I-00161, Roma, Italy}
\begin{abstract}
We investigate Anderson localization in three-dimensional disordered systems by comparing scalar classical waves with mass and force-constant disorder to electronic tight-binding models with diagonal and off-diagonal disorder. We show that the commonly employed mapping between classical-wave localization and the electronic Anderson model with diagonal disorder is not mathematically justified. Instead, the correct modulus-type formulation reveals that classical-wave systems constitute a distinct constrained disorder class, in which the acoustic sum rule correlates diagonal and off-diagonal matrix elements and prevents any direct correspondence with the standard electronic disorder models. Within a unified eigenvalue framework, we determine localization phase diagrams for all four disorder classes using complementary spectral, eigenvector, and level-statistics diagnostics. We find that classical-wave systems share a key qualitative feature with electronic off-diagonal disorder: localized states occur only near a band edge, while extended states persist in the central part of the spectrum even at strong disorder. At the same time, the acoustic sum rule produces localization topologies that differ fundamentally from both diagonal- and off-diagonal-disorder electronic systems. In particular, for mass disorder we obtain a phase diagram that differs qualitatively from previous results based on the conventional potential-type approach and reveals an extended localized regime near the upper band edge. Our results establish a unified perspective on localization in quantum and classical wave systems and provide new insight into the conditions under which Anderson localization may occur in three-dimensional photonic and acoustic media.
\end{abstract}
\date{\today}
\maketitle
\section{Introduction}
\label{sect:intro}
Since Anderson's seminal 1958 paper on the "absence of diffusion in certain random lattices"~\cite{anderson58,50years}, the localization of waves in disordered media has become one of the central problems of condensed-matter physics. Anderson demonstrated that electrons moving in a spatially fluctuating potential may cease to diffuse and instead become confined within a finite spatial region characterized by a localization length, $\xi$. Mott subsequently pointed out~\cite{mott67,mott87} that localized and extended states may coexist within the same spectrum and be separated by a {\it mobility edge}, thereby establishing the modern picture of the localization--delocalization transition.

A major advance in the understanding of this phenomenon came from the scaling theory of localization. Abrahams {\it et al.}~\cite{abrahams79} showed that Anderson localization originates from interference between time-reversed multiple-scattering paths, leading to enhanced back-scattering and the suppression of diffusive transport~\cite{lr85,chakravarty86}. Wegner further demonstrated~\cite{weg79} that the Anderson transition belongs to the universality class of the non-linear sigma model of planar ferromagnetism, placing localization within the broader framework of critical phenomena. These developments established that all states are localized in one and two dimensions, whereas a genuine localization--delocalization transition may occur in three dimensions. The Anderson problem was subsequently reformulated as an effective field theory with the action of a generalized non-linear sigma model~\cite{schafer80,mckane81}. An alternative field-theoretical description was developed by W\"olfle and Vollhardt~\cite{vollhardt80,vollhardt80a,vollhardt82}, whose self-consistent theory of localization was later interpreted in terms of stationary states in an effective potential well, the so-called potential-well analogy~\cite{economou83}.

Another highly successful approach emerged from the study of {\it local} spectral properties. Building on observations already contained in Anderson's original paper~\cite{anderson58}, it was realized~\cite{dobro97,dobro03,dobro10,byczuk05,byczuk10,schubert05,schubert10} that the geometrical average of the local density of states, the {\it typical} density of states (TDOS), may serve as an order parameter for the localization transition. In contrast to the arithmetic DOS, the TDOS vanishes whenever the local DOS becomes zero at any position in space and is therefore directly sensitive to localization. This framework also enabled the inclusion of electron--electron interactions and led to a unified description of Mott and Anderson localization mechanisms~\cite{dobro10,byczuk10}.

Because Anderson localization is essentially a wave-interference phenomenon, it was natural to ask whether the concepts developed for electrons could be extended to classical waves~\cite{john83,john83a,john84,john87}. The experimental observation of coherent back-scattering of light~\cite{wolf85}, often regarded as the optical analogue of weak localization, stimulated extensive efforts to observe strong localization of photons~\cite{anderson85,wiersma97,lagendijk96,sheng06,lagendijk09,genack11}. It was subsequently recognized~\cite{abdullaev80,deraedt89} that reduced dimensionality strongly enhances localization effects, motivating numerous studies of one- and two-dimensional photonic systems~\cite{genack11,schwartz07,karbasi12,innocenti15}.

The situation in three dimensions is considerably more subtle. Despite several experimental claims and investigations~\cite{wiersma97,wiersma97,storzer06,sperling12,scheffold13,sperling16,skipetrov16}, unambiguous evidence for Anderson localization of light in fully three-dimensional disordered media remains scarce. One possible reason is that localization in classical-wave systems often occupies only a limited spectral region, typically near a band edge~\cite{schirm98,amir13} or in the vicinity of a photonic band gap~\cite{john87}. As we shall discuss below, this behavior closely resembles that observed in electronic systems with off-diagonal disorder. A further complication arises from the existence of evanescent transport channels mediated by random longitudinal fields~\cite{nieuwenhuizen94,skipetrov14,skipetrov21}, which can obscure localization effects in three-dimensional dielectric systems.

Numerical studies have generally provided a clearer picture. Evidence for localization in three-dimensional classical-wave systems has been reported in a variety of settings, including hyper-uniform structures~\cite{haberko20,scheffold22} and media containing randomly distributed perfectly conducting obstacles~\cite{yamilov25}. More recently, near-field microscopic measurements on hyper-uniform photonic structures have also been interpreted as possible signatures of three-dimensional Anderson localization~\cite{granchi23}. These developments suggest that localization of classical waves in three dimensions is possible, but that its spectral location and physical origin may differ significantly from those of the conventional electronic Anderson problem.

On the theoretical side, however, the relationship between localization of classical waves and localization of electrons remains less clear than is often assumed. A large fraction of the existing literature is based on a mapping between the wave equation of a disordered classical medium and the Schr\"odinger equation of an electron moving in a random potential. While this correspondence has proved extremely influential and has motivated many important developments, its mathematical status deserves closer examination.

Indeed, in order to transfer the concepts successfully developed for the study of electronic localization to classical waves, it was proposed to establish a mapping between the classical acoustic-wave equation with spatially fluctuating mass density (or dielectric permittivity) and the Schr\"odinger equation with a disordered potential. For a scalar acoustic wave propagating in a medium with spatially disordered mass density, $m(\rr)=m_0+\Delta m(\rr)$, the frequency-domain wave amplitude $\phi(\rr,\omega)$ obeys
\be
\label{e1}
-\omega^2m(\rr) \phi(\rr,\omega) = K \nabla^2\phi(\rr,\omega),
\ee
where $K$ is the elastic constant. With the definitions
\be
E=\frac{m_0}{K}\omega^2,
\ee
and
\be
\label{pot2}
\VV(\rr,\omega)=-{\ts\frac{1}{K}}\omega^2\Delta m(\rr),
\ee
Eq.~(\ref{e1}) becomes
\be
\label{pot1}
E \phi(\rr,\omega)=-\nabla^2\phi(\rr,\omega)+\VV(\rr,\omega) \phi(\rr,\omega).
\ee
This equation has the formal appearance of a Schr\"odinger equation for the wave function $\phi(\rr,\omega)$ in the presence of a random potential $\VV(\rr,\omega)$. Building on this observation, John, Sompolinsky and Stephen~\cite{john83} applied the field-theoretical machinery originally developed for the Anderson problem~\cite{schafer80,mckane81} to classical acoustic waves, thereby establishing a conceptual bridge between quantum and classical localization.

The procedure embodied in Eqs.~(\ref{pot2}) and~(\ref{pot1}), referred to as the potential-type (PT) approach in Ref.~\cite{schirm18}, was subsequently generalized to Maxwell's equations~\cite{john84,john87}, where the mass density is replaced by a spatially fluctuating dielectric permittivity. In this form, the PT picture became the theoretical basis for numerous studies of localization in disordered optical and acoustic media~\cite{kroha93,kroha93,lagendijk96,sheng06,pinski12,pinski12a,vynck23}.

A closer inspection reveals, however, that the correspondence is not exact. The effective disorder potential introduced in Eq.~(\ref{pot2}) depends explicitly on the eigenvalue through the factor $\omega^2$. Consequently, the disorder term cannot be regarded as an eigenvalue-independent perturbation of a linear operator. The operator whose spectrum is sought therefore depends on the spectrum itself, a situation fundamentally different from the conventional Anderson Hamiltonian. From a mathematical perspective, the PT construction does not establish a rigorous mapping between the classical-wave problem and the electronic localization problem.

The correct formulation of the classical-wave eigenvalue problem was developed in~\cite{dyson53,maradudin68,monthus2010anderson,schirm24}. Instead of introducing an effective disorder potential, one first divides the equation of motion by the local mass density,
\be\label{MT}
E \phi(\rr,\omega)= \frac{m_0}{m(\rr)}\nabla^2\phi(\rr,\omega).
\ee
The disorder then enters through the inverse mass density rather than through an effective potential. An analogous construction applies to electromagnetic waves. In that case, the appropriate eigenvalue problem is obtained by dividing the wave equation by the dielectric permittivity~\cite{schirm24}. Equivalent formulations were also derived from the wave equation for the magnetic field~\cite{joannopoulos08,rotter17,schirm18,schirm24} or from the vector potential~\cite{viviescas03,zhu21}. In all cases, the inverse permittivity, {\it i.e.} the electric modulus, appears as part of the linear operator. For this reason, Ref.~\cite{schirm18} termed this formulation the modulus-type (MT) approach.

The distinction between the PT and MT formulations is not merely formal. In the PT picture, disorder fluctuations are multiplied by the square of the frequency, leading to predictions of a strong frequency dependence of the localization length~\cite{deraedt89,karbasi12a}. Such behavior is neither observed experimentally nor obtained within the MT framework~\cite{schirm18}. More importantly, the PT and MT approaches correspond to genuinely different eigenvalue problems and therefore need not produce identical localization phase diagrams.

Once the classical-wave problem is formulated correctly, the nature of its electronic analogue also changes. The MT description no longer resembles an electron moving in a fluctuating potential, corresponding to diagonal disorder. Instead, as discussed in the Appendix and in~\cite{levyleblond95}, it becomes closely related to an electronic problem with a spatially fluctuating effective mass, corresponding to off-diagonal disorder. This observation immediately raises a fundamental question: to what extent do the localization properties of classical waves and electronic systems with off-diagonal disorder actually coincide?

This question is particularly relevant because electronic off-diagonal disorder exhibits localization properties that differ qualitatively from those of the standard Anderson model. Earlier studies~\cite{economou77,antoniou77,weaire77} showed that states near the band center remain extended, whereas localization is confined to relatively narrow spectral regions near the band edges. In contrast, electronic systems with diagonal disorder undergo a conventional Anderson transition in which localization progressively invades the spectrum as disorder increases. Determining which of these scenarios is realized in classical-wave systems is therefore a central issue in understanding localization in disordered media.

The present work addresses this question by comparing localization in four representative disorder classes: electronic tight-binding models with diagonal disorder (DD) and off-diagonal disorder (ODD), and scalar classical-wave systems with mass disorder (MD) and force-constant disorder (FCD). Our objective is not merely to compare localization phase diagrams, but to establish which aspects of localization are generic and which arise from structural constraints specific to the underlying wave equation.

A crucial difference between the electronic and classical problems is that the latter obeys the acoustic sum rule. In a classical dynamical matrix, diagonal and off-diagonal matrix elements are not independent but are linked by a conservation law expressing translational invariance. This correlation has no counterpart in the standard Anderson models, where diagonal and hopping disorder can be assigned independently. As a consequence, classical-wave systems do not simply realize either diagonal or off-diagonal electronic disorder. Instead, they define a distinct constrained disorder class whose localization properties must be established independently.

This observation has important implications. On the one hand, the MT formulation suggests a close connection between classical waves and electronic systems with off-diagonal disorder. On the other hand, the acoustic sum rule implies that any such correspondence can only be partial. Similarities in localization behavior therefore do not necessarily imply a genuine mapping between the underlying disorder classes. Determining the extent of these similarities and differences is one of the principal goals of the present work.

To address this issue, we formulate all four disorder classes within a common eigenvalue framework and analyze them using identical localization diagnostics. In particular, we determine mobility edges from the participation ratio, the typical density of states, and level-spacing statistics, and use these observables to construct localization phase diagrams in the disorder--eigenvalue plane. This unified approach allows a direct comparison of the spectral topology of localized and extended states across both quantum and classical systems.

Our analysis reveals that classical-wave systems share an important qualitative feature with electronic off-diagonal disorder: localized states are confined to spectral regions near a band edge. At the same time, the acoustic sum rule generates correlations between diagonal and off-diagonal matrix elements that fundamentally distinguish the classical problem from both DD and ODD electronic systems. We therefore find that no direct mapping exists between the standard electronic Anderson models and localization of classical waves, despite the similarities observed in their phase diagrams.

A second central result concerns the mass-disordered classical-wave problem. We show that the phase diagram obtained within the correct MT formulation differs qualitatively from that predicted by the PT approach. In particular, the MT description reveals an extended high-frequency localized sector that has no counterpart in the PT framework. The discrepancy reflects not a quantitative correction but the fact that the two approaches correspond to different eigenvalue problems.

The remainder of the paper is organized as follows. We first establish a common formalism for the electronic and classical systems and identify the structural role of the acoustic sum rule. We then determine and compare the localization phase diagrams of the four disorder classes using complementary localization diagnostics. Finally, we discuss the implications of our findings for the long-standing search for Anderson localization of light and other classical waves in three-dimensional disordered media.
\section{General Formalism}
The purpose of this section is to establish a common framework for comparing localization in quantum and classical disordered systems while making explicit the role of the acoustic sum rule. Our analysis proceeds along three complementary directions. First, we formulate all four disorder classes considered in this work—force-constant disorder and mass disorder for classical waves, and diagonal and off-diagonal disorder for electrons—as symmetric eigenvalue problems. This unified representation makes transparent both the similarities and the fundamental differences between the quantum and classical cases. Second, we compute, for each model and over a broad range of disorder strengths, a common set of spectral and eigenvector-based observables, including the arithmetic and typical densities of states, the participation ratio, and level statistics based on adjacent-gap ratios. Third, we use these observables to determine mobility edges and construct localization phase diagrams that can be directly compared across all disorder classes.

A central outcome of this approach is the determination of a new localization phase diagram for classical waves with mass disorder within the correct modulus-type (MT) formulation. As we shall show, this phase diagram differs qualitatively from the previously published results obtained within the potential-type (PT) framework.

The unified framework also allows us to examine critically the proposal~\cite{dobro97,dobro03,dobro10,byczuk05,byczuk10,schubert05,schubert10} that the geometrically averaged local density of states acts as an order parameter for Anderson localization. In particular, we test whether the support of the typical DOS coincides with the region of extended states identified independently from eigenvector-based diagnostics and level-statistics analyses. The availability of several complementary localization measures makes it possible to assess the robustness of this identification across different disorder classes.

A key conceptual ingredient underlying the present comparison is that all four systems can be cast into the same mathematical form, yet belong to different disorder classes. In electronic systems, diagonal and hopping disorder can be specified independently, generating a broad family of random matrices within a given symmetry class. In contrast, classical-wave systems obey the acoustic sum rule, which imposes a conservation law linking diagonal and off-diagonal matrix elements. As a consequence, disorder in the classical problem is intrinsically constrained. This distinction plays a central role throughout the paper and ultimately prevents a direct mapping between the standard electronic Anderson models and the localization problem of classical waves.
\begin{table}[t]
\centering
\begin{tabular*}{\columnwidth}{@{\extracolsep{\fill}}ccccc}
\hline \hline
& MD & FCD & DD & ODD\\
\hline
&&&&\\
$H_{ii}$  & $\sum_\ell \dfrac{K_0}{\sqrt{m_i m_\ell}}$ & $\sum_\ell \dfrac{1}{m_0}K_{i\ell}$ & $\epsilon_i$ & $\epsilon_0$ \\
$H_{ij}$  & $-\dfrac{K_0}{\sqrt{m_i m_j}}$             & $-\dfrac{1}{m_0}K_{ij}$             & $V_0$        & $V_{ij}$     \\
$\lambda$ & $\omega^2$                                 & $\omega^2$                          & $E$          & $E$          \\
$\phi_i$  & $\sqrt{m_i}u_i$                            & $u_i$                               & $\psi_i$     & $\psi_i$     \\
&&&&\\
\hline \hline
\end{tabular*}
\caption{Correspondence between the matrix elements entering the unified eigenvalue problem, Eq.~(\ref{eq:unified_evp}), for the four disorder classes studied in this work: mass disorder (MD), force-constant disorder (FCD), diagonal disorder (DD), and off-diagonal disorder (ODD). Although all four systems can be expressed within the same eigenvalue framework, the table makes explicit how disorder enters the underlying operators, and foreshadows the key distinction between unconstrained electronic disorder and the constrained disorder classes generated by the acoustic sum rule in classical-wave systems.}
\label{table1}
\end{table}

From a computational perspective, the unified formulation offers an important practical advantage. Because all observables are derived from the full spectrum and eigenvectors obtained by exact diagonalisation, spectral and eigenvector-based diagnostics can be evaluated on an equal footing. In particular, the local density of states can be constructed directly from the eigenmodes, allowing both arithmetic and geometric averages to be determined without additional approximations.

The results presented below demonstrate that, despite the diversity of microscopic realizations, localization in all four models can be analyzed within a common framework. At the same time, systematic differences emerge that reflect the presence or absence of conservation laws, the way disorder enters the eigenvalue problem, and the structure of the underlying spectrum. These similarities and differences become particularly transparent in the localization phase diagrams, which provide a global view of the distribution of localized and extended states. In particular, applying the modulus-type approach~\cite{dyson53,maradudin68,monthus2010anderson}, we find that three-dimensional classical waves with mass disorder, much like the force-constant-disorder case, exhibit localized states only in a spectral region near the upper band edge. More importantly, we find that neither classical disorder class can be mapped directly onto the standard electronic diagonal- or off-diagonal-disorder problems.
\subsection{Unified eigenvalue formulation}
The four models considered in this work can all be written as a real-symmetric eigenvalue problem of the form,
\begin{equation}
\sum_j H_{ij}\phi_j=\lambda \phi_i,
\label{eq:unified_evp}
\end{equation}
where the matrix $H$ is real symmetric, $H_{ij}=H_{ji}$, and therefore Hermitian. The physical meaning of $H_{ij}$, $\lambda$, and $\phi_i$ depends on the system under consideration. Table~\ref{table1} provides the explicit correspondence. For the electronic models, $\lambda$ denotes the energy $E$ and $\phi_i$ is the electronic wave function $\psi_i$. For the vibrational models, $\lambda$ corresponds to the squared frequency, $\omega^2$, while $\phi_i$ represents either the displacement field itself or a mass-weighted displacement, depending on the type of disorder.

Eq.~(\ref{eq:unified_evp}) provides the natural starting point for a unified discussion of localization. Once expressed in this form, all four systems admit the same set of localization diagnostics, including the density of states, local density of states, participation ratio, and level statistics. At the same time, the relations between diagonal and off-diagonal matrix elements encode the physical constraints specific to each problem. It is precisely at this level that the essential distinction between classical-wave and electronic localization emerges.
\subsection{Electronic tight-binding models}
For the electronic problem, we consider the canonical nearest-neighbor tight-binding (TB) equation in site representation,
\begin{equation}
E \psi_i = \epsilon_i \psi_i + \sum_{j\in \partial i} V_{ij}\psi_j,
\label{eq:tb_general}
\end{equation}
where $\epsilon_i$ denotes the on-site energy, $V_{ij}$ the hopping amplitude between neighboring sites, and $\partial i$ the set of nearest neighbors of site $i$. As summarized in Appendix B, this form of the Anderson model can be obtained by discretizing a continuum Schr\"odinger equation with spatially fluctuating potentials and/or effective masses on a simple-cubic lattice.

Eq.~(\ref{eq:tb_general}) can be written directly in the unified form of Eq.~(\ref{eq:unified_evp}), with
\begin{equation}
H_{ij} = \epsilon_i\delta_{ij} + V_{ij},\quad\lambda=E,\quad \phi_i=\psi_i.
\end{equation}
The two electronic disorder classes considered in this work correspond to different choices for the random matrix elements.
\paragraph{Diagonal disorder (DD).}
In the diagonal-disorder case, randomness acts exclusively on the on-site energies,
\begin{equation}
\epsilon_i \in [\epsilon_0-\Delta\epsilon/2,\epsilon_0+\Delta\epsilon/2],
\end{equation}
while the hopping amplitudes remain fixed, $V_{ij}=V_0$. Throughout this work we choose $\epsilon_0=0$ and $V_0=1$. Disorder therefore resides entirely in the diagonal sector of the Hamiltonian, whereas the off-diagonal matrix elements merely encode the connectivity of the underlying lattice. This is the standard Anderson model and has been extensively studied numerically~\cite{bulka1987localization,grussbach95,dobro03,schubert05,schubert10}.
\paragraph{Off-diagonal disorder (ODD).}
In the off-diagonal-disorder case, the on-site energy is uniform, $\epsilon_i=\epsilon_0$, while the hopping amplitudes fluctuate according to
\begin{equation}
V_{ij}\in [V_0-\Delta V/2,V_0+\Delta V/2].
\end{equation}
Here we again choose $\epsilon_0=0$ and $V_0=1$. Disorder is therefore confined to the off-diagonal matrix elements of the Hamiltonian, while the diagonal sector remains unchanged. This model has been investigated in a number of studies~\cite{economou77,antoniou77,weaire77}.

A key feature of both electronic disorder classes is that diagonal and off-diagonal matrix elements can be specified independently. In particular, there is no constraint relating $H_{ii}$ to the neighboring hopping amplitudes $H_{ij}$. Unlike the classical-wave systems discussed below, electronic tight-binding models are not subject to a conservation law imposing a sum rule between diagonal and off-diagonal sectors. As a result, modifying the diagonal disorder or the hopping disorder changes the spectrum without any compensating structural constraint. Within the symmetry class defined by Hermiticity, the electronic Hamiltonian therefore represents an unconstrained random-matrix problem.

As we shall see below, this independence of diagonal and off-diagonal disorder constitutes one of the most important distinctions between the electronic and classical-wave localization problems. The acoustic sum rule correlates the corresponding matrix elements in the classical case, generating a constrained disorder class whose localization properties differ fundamentally from those of both DD and ODD electronic systems.
\subsection{Scalar classical-wave models}
For consistency with the electronic case, we restrict ourselves to schematically discretized classical-wave models with scalar wave amplitudes $u_i(t)=u(\rr_i,t)$. The consequences of the vector nature of elastic and electromagnetic waves in disordered media have been discussed elsewhere~\cite{schirm14,skipetrov21}. The present scalar formulation retains the essential ingredients needed to analyze localization and, at the same time, permits a direct comparison with the electronic tight-binding models introduced above.

Let $u_i$ denote the scalar displacement at site $i$, $m_i$ the local mass, and $K_{ij}=K_{ji}$ the spring constant coupling nearest-neighbor sites $i$ and $j$. The equations of motion are
\begin{equation}
m_i \ddot{u}_i = - \sum_{j\in \partial i} K_{ij}(u_i-u_j).
\label{eq:eom_phonon}
\end{equation}
Searching for harmonic solutions of the form $u_i(t)=u_i e^{-i\omega t}$ yields
\begin{equation}
\omega^2 m_i u_i = \sum_{j\in \partial i} K_{ij}(u_i-u_j).
\label{eq:phonon_raw}
\end{equation}
As shown in Appendix A, this equation can be obtained by discretizing a continuous scalar-wave equation with spatially fluctuating elastic modulus or mass density. Depending on the type of disorder, different representations are most convenient for casting Eq.~(\ref{eq:phonon_raw}) into the unified eigenvalue form of Eq.~(\ref{eq:unified_evp}).
\paragraph{Mass disorder (MD).}
When the spring constants are uniform, $K_{ij}=K_0$, and the masses fluctuate according to
\begin{equation}
m_i \in [m_0-\Delta m/2,m_0+\Delta m/2],
\label{eq:mass-distr}
\end{equation}
Eq.~(\ref{eq:phonon_raw}) reduces to
\begin{equation}\label{e14}
\omega^2 m_i u_i = \sum_{j\in \partial i} K_0(u_i-u_j).
\end{equation}
The discretized version of the previously employed PT approach
\cite{pinski12,pinski12a}
consists in writing $m_i=m_0+\Delta m_i$ and recasting Eq.~(\ref{e14}) as 
\be\label{discrete-pt}
(\underbrace{6K_0-\omega^2m_0}_{E})u_i=\underbrace{\Delta m_i\,\omega^2}_{\epsilon_i(\omega)}u_i+K_0\sum_{j\in\partial i}u_j 
\ee
which formally maps the MD problem onto the electronic DD problem, albeit with an effective on-site energy that depends explicitly on the eigenvalue. Eq.~(\ref{discrete-pt}) has indeed been used to analyze mass disorder by establishing a correspondence with diagonal electronic disorder~\cite{pinski12,pinski12a}. As discussed above and in~\cite{schirm18,schirm24}, this procedure is mathematically problematic because it transfers part of the eigenvalue term into the Hamiltonian whose spectrum is being determined. The PT construction therefore does not merely provide an alternative representation of the same problem; it defines a different eigenvalue problem. For this reason, we do {\it not} adopt the PT formulation and instead employ the correct modulus-type (MT) approach. As we shall demonstrate below, the two formulations lead to qualitatively different localization phase diagrams.

After dividing Eq.~(\ref{e14}) by $m_i$, the resulting operator is no longer Hermitian. One therefore introduces the mass-weighted variable~\cite{dyson53,maradudin68,monthus2010anderson,schirm24}
\begin{equation}\label{massweight}
\phi_i = \sqrt{m_i} u_i.
\end{equation}
The eigenvalue problem then becomes symmetric, with matrix elements
\begin{equation}
H_{ii}=\sum_{\ell\in \partial i}\frac{K_0}{\sqrt{m_i m_\ell}},
\quad H_{ij}=-\frac{K_0}{\sqrt{m_i m_j}} \quad (i\neq j),
\label{eq:md_matrix}
\end{equation}
and eigenvalue $\lambda=\omega^2$. This corresponds to the MD column of Table~\ref{table1}. Throughout this work we set $m_0=K_0=1$. The mass-disorder problem has been investigated previously in~\cite{monthus2010anderson,chaudhuri2010heat,pinski12,pinski12a}, among others. Unlike~\cite{pinski12,pinski12a}, which employed the PT formulation, we do not consider negative masses. Consequently, the physically accessible disorder range is restricted to $\Delta m<2$.
\paragraph{Force-constant disorder (FCD).}
When the masses are uniform, $m_i=m_0$, while the spring constants fluctuate according to
\begin{equation}
K_{ij}\in [K_0-\Delta k/2,K_0+\Delta k/2],
\end{equation}
Eq.~(\ref{eq:phonon_raw}) becomes
\begin{equation}
\omega^2 m_0 \,u_i = \sum_{j\in \partial i} K_{ij}(u_i-u_j).
\label{eq:fcd_dyn}
\end{equation}
In this case the displacement itself is already the natural eigenvector variable, $\phi_i=u_i$, and no additional transformation is required. The corresponding symmetric matrix elements are
\begin{equation}
H_{ii}=\sum_{\ell\in \partial i}\frac{K_{i\ell}}{m_0},
\quad
H_{ij}=-\frac{K_{ij}}{m_0}
\quad (i\neq j),
\label{eq:fcd_matrix}
\end{equation}
with $\lambda=\omega^2$. This corresponds to the FCD column of Table~\ref{table1}. As in the mass-disorder case, we choose $m_0=K_0=1$ throughout.
\subsection{Presence/absence of conservation rule for classical waves and electrons}
The formal analogy summarized in Table~\ref{table1} should not obscure the most important physical distinction between classical waves and electrons: in vibrational systems the matrix elements are not independent. Instead, they obey a conservation law that reflects the translational invariance of the underlying elastic medium. Indeed, Eq.~\eqref{eq:phonon_raw} depends only on displacement differences $(u_i-u_j)$, and a uniform translation, $u_i=u_0\;\forall i$, produces no restoring force. The existence of such a zero-frequency mode implies the sum rule
\begin{equation}
\sum_j H_{ij}=0,
\label{eq:sum_rule}
\end{equation}
which fixes the diagonal matrix elements according to
\begin{equation}\label{sumrule}
H_{ii} = - \sum_{j\neq i} H_{ij}.
\end{equation}

For the FCD model this relation follows directly from Eq.~\eqref{eq:fcd_matrix}. The diagonal element at site $i$ is completely determined by the spring constants connecting that site to its neighbors and therefore does not constitute an independent degree of freedom. Disorder in the off-diagonal couplings automatically generates correlated disorder in the diagonal sector.

The same principle applies to the MD model after the transformation to mass-weighted coordinates. Inspection of Eq.~(\ref{eq:md_matrix}) shows that diagonal and off-diagonal matrix elements remain linked through the local mass environment. Although the explicit form of the matrix differs from the FCD case, the physical origin of the constraint is identical: a spatially uniform displacement field must remain a zero-cost mode. The acoustic sum rule therefore survives the transformation to a symmetric eigenvalue problem and continues to constrain the structure of the disorder.

This conservation law has several important consequences for localization. First, it protects the existence of low-frequency acoustic modes, which remain extended. Indeed, previous studies of force-constant disorder~\cite{amir13} show that even at strong disorder a finite spectral region of delocalized plane-wave-like states persists near $\omega=0$. Second, the sum rule strongly correlates diagonal and off-diagonal matrix elements, making the classical-wave problem qualitatively different from a generic random-matrix ensemble or from a standard electronic tight-binding Hamiltonian with independently assigned site and bond disorder. Third, localization in classical-wave systems does not originate from arbitrary diagonal randomness but from disorder subject to a mechanical compatibility constraint imposed by translational invariance.

No analogous sum rule exists for the electronic models considered here. In the tight-binding Hamiltonian of Eq.~(\ref{eq:tb_general}), the on-site energies $\epsilon_i$ and hopping amplitudes $V_{ij}$ are independent microscopic parameters. Only in the special case of a disordered electronic system whose disorder originates exclusively from a spatially varying effective mass~\cite{levyleblond95}, rather than from a fluctuating potential, does a related conservation law emerge. As discussed in Appendix B, this constraint is associated with the existence of low-energy free-electron states.
\subsection{Physical consequences for localization}
The common eigenvalue structure of Eq.~(\ref{eq:unified_evp}) explains why both electrons and classical waves can exhibit Anderson localization, while the conservation law expressed by Eq.~(\ref{eq:sum_rule}) clarifies why their localization phase diagrams are not identical. The shared mathematical framework allows localization phenomena in all four systems to be analyzed using the same diagnostics, but the constraints imposed on the matrix elements lead to qualitatively different spectral topologies.

In the standard electronic DD model, localized states emerge at the band edges as disorder increases and progressively invade the spectrum until, beyond a critical disorder strength, all states become localized. The situation is markedly different in the electronic ODD model, where localization remains confined to spectral regions near the band edges while states near the band center stay extended. As we shall show below, classical-wave systems display behavior that is much closer to the latter scenario.

In classical-wave systems, the acoustic sum rule constrains the low-frequency sector and protects long-wavelength modes from localization. As a consequence, localization develops predominantly at high frequencies, while extended acoustic excitations survive near $\omega=0$. In the FCD case, sufficiently strong disorder may even generate negative values of $\omega^2$, signaling mechanical instability. This instability is closely related to the unstable normal-mode spectra encountered in liquids~\cite{taraskin02,schirm22Modeling,mossa23}. Within this interpretation, the instantaneous-normal-mode spectrum of a liquid~\cite{keyes97} can be viewed as that of an FCD-type disordered elastic medium whose disorder strength is controlled by temperature.

The MD case can be interpreted from a similar perspective. After the mass-weighting transformation of Eq.~(\ref{massweight}), the problem is effectively mapped onto a particular class of force-constant disorder. Unlike the generic FCD case, however, stability is preserved because the masses are restricted to remain positive.

In all four disorder classes considered here, localization is ultimately determined by the spectrum and eigenvectors of the matrix $H$. This common structure makes it possible to compare electrons and classical waves on equal footing and to construct localization phase diagrams within a unified framework. At the same time, the present analysis demonstrates that the decisive distinction is not simply whether disorder enters diagonal or off-diagonal matrix elements, but whether these matrix elements are independent or constrained.

From this perspective, both FCD and MD should be viewed as constrained disorder classes. The acoustic sum rule induces correlated fluctuations in the diagonal and off-diagonal sectors and thereby generates a type of disorder that has no direct counterpart in the standard electronic Anderson models. Classical-wave localization therefore shares important qualitative features with electronic off-diagonal disorder while remaining fundamentally distinct because of the underlying conservation law.
\subsection{Observables and mobility-edge criteria}
We solve numerically the eigenvalue problem of Eq.~(\ref{eq:unified_evp}) for all disorder classes listed in Table~\ref{table1}. All calculations were performed using JAX~\cite{jax2018github} for random-matrix generation, data handling, and diagonalisation. The eigenvalue problems were solved by full diagonalisation using dense eigensolvers (LAPACK routines accessed through \texttt{jax.scipy.linalg.eigh}), providing direct access to the complete spectrum, ${\lambda_n}$, and the corresponding eigenvectors, ${\phi_i^{(n)}}$. We considered simple-cubic lattices with $N=L^3$ sites and linear sizes $L=$~10, 13, 16, 20, and 26 ($N=$~1000, 2197, 4096, 8000, and 17576). Each lattice site has coordination number $\mathrm{z}=$~6. Disorder averages were performed over 800, 400, 200, 100, and 50 independent realizations, respectively.

To characterize localization we employ three complementary diagnostics: the typical density of states, the participation ratio, and adjacent-gap level statistics. Together, these quantities provide independent information on spectral support, eigenvector structure, and eigenvalue correlations, allowing mobility edges to be identified with high confidence.
\subsubsection{Typical density of states}
The local density of states projected onto site $i$ is defined as~\cite{schubert05}
\be
\rho_i(\lambda)={\sum_{n=1}^N |\phi_i^{(n)}|^2\,\delta(\lambda-\lambda_n)},
\label{eq:ldos_num}
\ee
from which we construct the arithmetic (arithmetically averaged) density of states and the typical density of states,
\be
\rho(\lambda)=\left\langle \frac{1}{N}\sum_{i=1}^N \rho_i(\lambda)\right\rangle
=\left\langle\sum_{n=1}^N\delta(\lambda-\lambda_n)\right\rangle,
\ee
\be
\rho_{\mathrm{typ}}(\lambda)=\exp\!\left[
\left\langle \frac{1}{N}\sum_{i=1}^N \ln\rho_i(\lambda)
\right\rangle
\right].
\label{eq:rhotyp_num}
\ee
The support of $\rho(\lambda)$ determines the spectral band edges, whereas the support of $\rho_{\mathrm{typ}}(\lambda)$ provides an estimate of the mobility edges. Throughout this work, $\langle\cdots\rangle$ denotes an average over disorder realizations. The typical DOS plays a central role because it is sensitive to the spatial fragmentation of eigenstates and is expected to vanish in localized spectral regions~\cite{dobro97,dobro03,dobro10,byczuk05,byczuk10,schubert05,schubert10}.
\subsubsection{Participation ratio}
As an independent eigenvector-based measure of localization we employ the participation ratio,
\begin{equation}
P(\lambda_n)=\frac{1}{\sum_{i=1}^N |\phi_i^{(n)}|^4},
\label{eq:pr_main}
\end{equation}
which estimates the effective number of sites occupied by an eigenmode. Its finite-size scaling distinguishes localized from extended states. For extended modes, $|\phi_i|^2\sim 1/N$, implying $P\sim N\sim L^3$, whereas for localized modes $P\sim \xi^3$ remains finite for $L\gg\xi$.

In three dimensions it is convenient to introduce the characteristic length scale
\begin{equation}
\xi_{\mathrm{PR}} \equiv P^{1/3},
\end{equation}
for which $\xi_{\mathrm{PR}}\sim L$ in the extended regime and $\xi_{\mathrm{PR}}\sim \xi$ in the localized regime. The dimensionless quantity
\begin{equation}
\frac{\xi_{\mathrm{PR}}}{L}=\frac{P^{1/3}}{L}
\end{equation}
therefore provides a useful finite-size-scaling observable. It approaches a constant for extended states and decreases as $\xi/L$ for localized states. At a mobility edge $\lambda_c$, where the localization length becomes comparable to the system size, scale invariance implies an approximately size-independent value of $\xi_{\mathrm{PR}}/L$.

As a consequence, curves of $\xi_{\mathrm{PR}}/L$ obtained for different system sizes exhibit a crossing near $\lambda_c$, providing a robust estimate of the mobility edge. More generally, the participation ratio obeys the scaling form
\begin{equation}
P(\lambda,L)=L^{D_2}\,f\!\left[ (\lambda-\lambda_c)L^{1/\nu}\right],
\end{equation}
where $D_2$ denotes the correlation (multi-fractal) dimension and $\nu$ the localization-length exponent. One has $D_2=3$ in the extended phase, $D_2=0$ in the localized phase, and $0<D_2<3$ at criticality, where the eigenstates are multi-fractal. The crossing of $P/L^{D_2}$ at $\lambda_c$ follows directly from this scale-invariant form.

In practice, we identify mobility edges from crossings of $P/N^{1/3}$, or equivalently $\xi_{\mathrm{PR}}/L$, as a function of eigenvalue. This criterion is particularly useful when $\rho_{\mathrm{typ}}(\lambda)$ becomes very small and numerically noisy, as frequently occurs near band edges and at strong disorder.
\subsubsection{Level-distance statistics}
A third, purely spectral diagnostic is provided by the adjacent-gap ratio
\be
r_n=\frac{\min(\delta_n,\delta_{n+1})}{\max(\delta_n,\delta_{n+1})},
\qquad
\delta_n=\lambda_{n+1}-\lambda_n.
\label{eq:average_r}
\ee
which depends only on the eigenvalues and not on the eigenvectors. Its disorder-averaged value, $\langle r\rangle$, distinguishes localized and extended regimes through the correlations between neighboring levels. In the extended phase, level repulsion leads to Wigner--Dyson statistics and $\langle r\rangle \simeq 0.536$ for the Gaussian orthogonal ensemble (GOE). In the localized phase, eigenvalues become statistically independent and Poisson statistics yields $\langle r\rangle \simeq 0.386$. These limiting values are well established~\cite{oganesyan07,atas13}.

Along any path in the disorder--eigenvalue plane, $\langle r\rangle$ exhibits a smooth crossover between the GOE and Poisson limits upon crossing a mobility edge. This crossover sharpens systematically with increasing system size. The adjacent-gap ratio therefore provides a robust and entirely independent indicator of localization, complementing both the participation ratio and the typical DOS. In the following, we use it as a decisive diagnostic whenever the mobility edges inferred from $\rho_{\mathrm{typ}}(\lambda)$ and the participation ratio show significant discrepancies.
\begin{figure*}[t]
\centering
\includegraphics[width=0.8\textwidth]{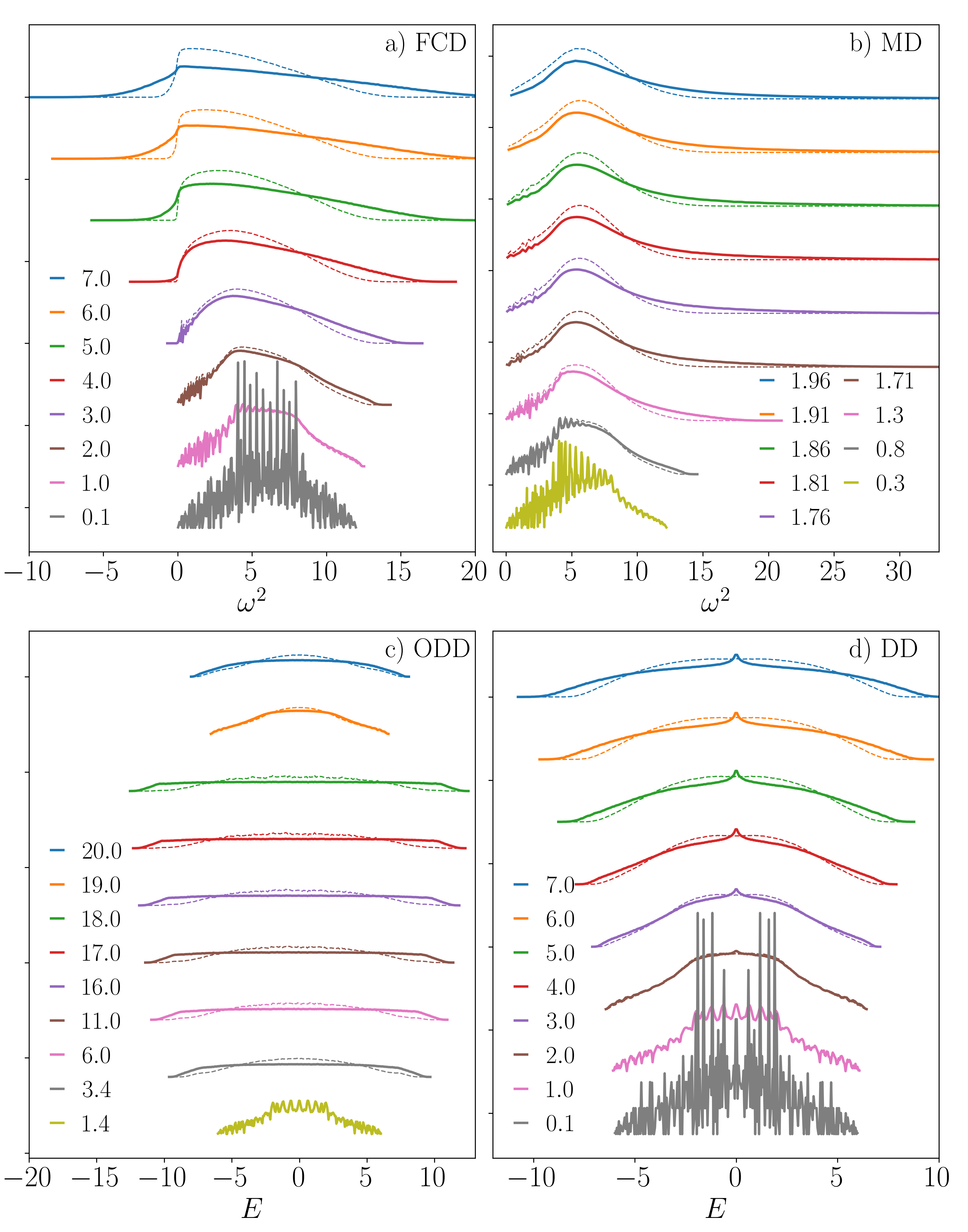}
\caption{Arithmetic (solid lines) and typical (dashed lines) densities of states for the four disorder classes introduced in Table \ref{table1}: a) classical waves with force-constant disorder (FCD); b) classical waves with mass disorder (MD); c) electrons with off-diagonal disorder (ODD); and  d) electrons with diagonal disorder (DD). Curves are shown for increasing disorder strengths $\Delta k$, $\Delta m$, $\Delta \epsilon$, $\Delta V$ (indicated in the legends), and are vertically shifted for clarity.}
\label{fig:dos_ldos_png}
\end{figure*}
\begin{figure*}[t]
\centering
\includegraphics[width=1\textwidth]{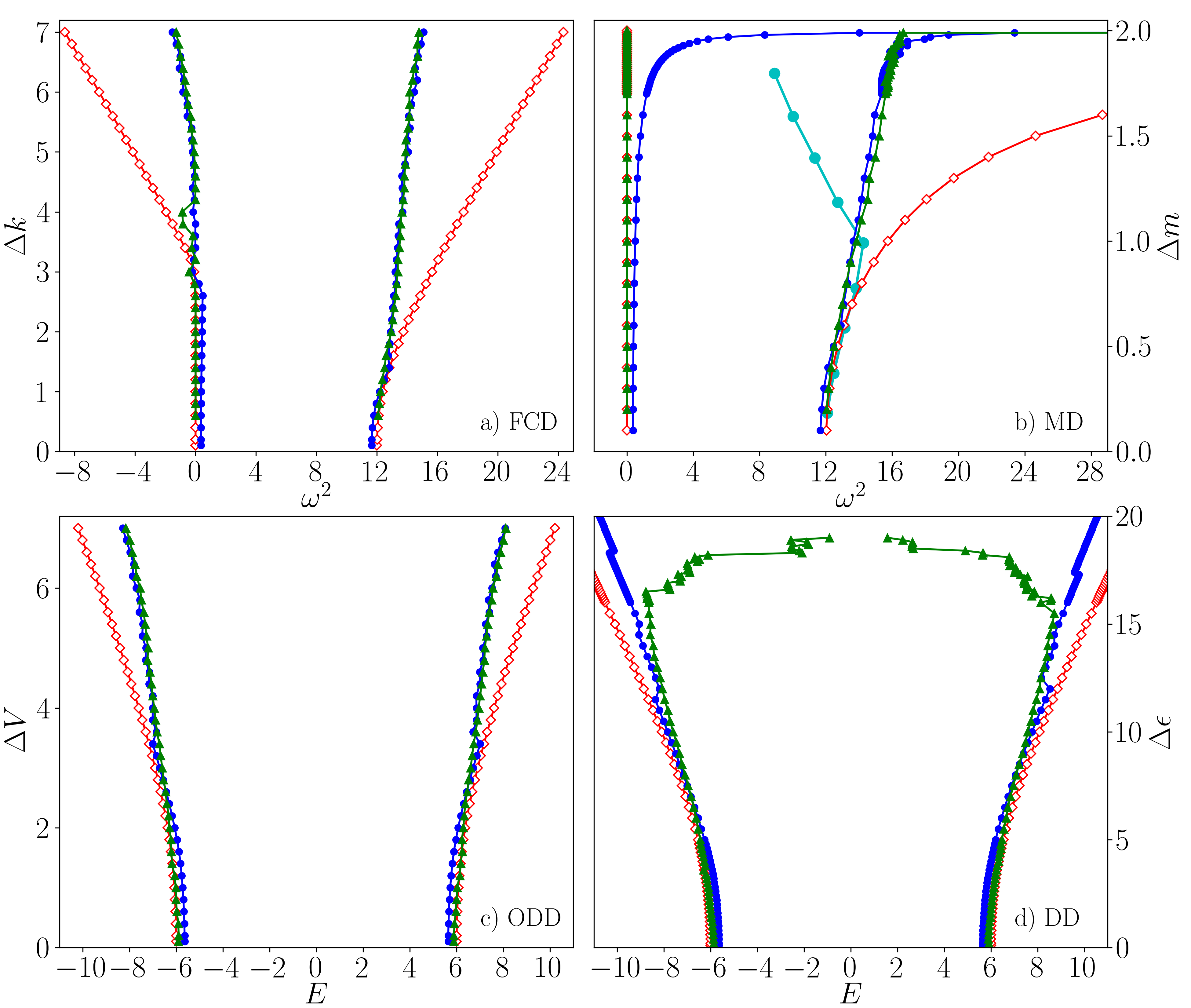}
\caption{Localization phase diagrams for classical-wave and electronic disordered systems: a) classical waves with force-constant disorder (FCD); b) classical waves with mass disorder (MD); c) electrons with off-diagonal disorder (ODD); and d) electrons with diagonal disorder (DD). Mobility edges separating extended and localized states are determined from the typical density of states $\rho_{\mathrm{typ}}$ (blue full circles) and the participation ratio $\mathcal{P}$ (green triangles), and compared with band edges (red open diamonds). The full cyan circles in panel b) are the mobility edges using the potential-type (PT) scheme of Eqs. (\ref{pot1}) and (\ref{pot2}), taken from~\cite{pinski12}. In the FCD case (panel a), large disorder induces unstable modes with $\omega^2<0$.}
\label{fig2}
\end{figure*}
\section{Results}
\label{sec:numerical results}
\subsection{Arithmetic versus typical DOS}
The evolution of the arithmetic and typical densities of states shown in Fig.~\ref{fig:dos_ldos_png} provides a first global view of the localization properties of the four disorder classes introduced above. In particular, the comparison between $\rho(\lambda)$ and $\rho_{\mathrm{typ}}(\lambda)$ reveals how localization develops across the spectrum and already foreshadows the different phase-diagram topologies discussed below.

At weak disorder, all systems exhibit a common feature: the spectrum consists of a collection of narrow peaks reflecting the discrete nature of the finite lattice. In this regime, the density of states is essentially a superposition of weakly perturbed eigenmodes of the corresponding ordered system. The persistence of these sharp structures, particularly visible in the lowest-disorder curves, indicates that disorder is still insufficient to induce substantial level mixing or spectral broadening. As the disorder strength increases, neighboring peaks progressively overlap and merge into smooth spectral distributions, signaling the onset of strong hybridization between eigenmodes.

A second feature common to all models is the progressive reduction of the support of $\rho_{\mathrm{typ}}(\lambda)$ relative to that of $\rho(\lambda)$ as disorder increases. Correspondingly, $\rho_{\mathrm{typ}}(\lambda)$ becomes increasingly suppressed near the spectral boundaries while the arithmetic DOS remains finite. Within the TDOS framework, these regions are naturally associated with localized states, whereas spectral regions where $\rho_{\mathrm{typ}}(\lambda)$ remains finite are identified with extended states.

For the classical-wave models, panels a) and b), the spectra are expressed in terms of the eigenvalue $\omega^2$ and are constrained by mechanical stability. In the force-constant-disorder (FCD) case, increasing $\Delta k$ produces a pronounced distortion of the spectrum, accompanied by a progressive transfer of spectral weight toward lower frequencies. At sufficiently strong disorder, fluctuations generate negative values of $\omega^2$, indicating the appearance of mechanically unstable modes. Simultaneously, $\rho_{\mathrm{typ}}(\lambda)$ becomes strongly suppressed over a broad spectral interval, reflecting both localization effects and the approach to mechanical instability.

An interesting feature of the FCD spectra is that the system remains mechanically stable up to $\Delta K=3$, despite the fact that the disorder distribution already includes negative force constants. This demonstrates that a force-constant network can tolerate a finite fraction of negative springs without becoming unstable~\cite{schirm98}. Mechanical stability is therefore controlled by collective network properties rather than by the sign of individual force constants.

The mass-disorder (MD) case, panel b), displays a qualitatively different evolution. Because the masses are restricted to remain positive ($\Delta m<2$), the spectrum remains positive definite and no unstable branch develops. The spectra reported here were obtained using the correct modulus-type (MT) formulation rather than the conventional potential-type (PT) construction of Eqs.~(\ref{pot2}) and~(\ref{pot1}). Since the PT and MT approaches correspond to different eigenvalue problems, it is not surprising that the resulting spectra differ from those reported by Pinsky \emph{et al.}~\cite{pinski12,pinski12a}, which were obtained within the PT framework~\footnote{Mass disorder has previously been treated using the correct MT formulation~\cite{monthus2010anderson}. However, neither the systematic evolution of the spectrum with increasing $\Delta m$ nor the corresponding localization phase diagram was reported.}.

With increasing disorder, the spectrum broadens but retains a comparatively regular shape. Most importantly, the suppression of $\rho_{\mathrm{typ}}(\lambda)$ occurs predominantly near the upper spectral edge, indicating that localization develops first in the high-frequency sector while low-frequency modes remain extended. This behavior already suggests a localization topology fundamentally different from that of the standard Anderson model and anticipates the phase diagrams discussed below.

The electronic systems, panels c) and d), exhibit a markedly different behavior. Because they are not constrained by the acoustic sum rule, their spectra are approximately symmetric with respect to the energy variable $E$, and localization develops from both spectral edges. This is reflected in the progressive suppression of $\rho_{\mathrm{typ}}(\lambda)$ in the band tails, while a finite typical DOS survives near the band center~\cite{antoniou77}.

For off-diagonal disorder (ODD), the central spectral region remains remarkably robust against localization even at strong disorder, consistent with earlier studies~\cite{economou77,antoniou77,weaire77}. The arithmetic DOS broadens substantially, yet $\rho_{\mathrm{typ}}(\lambda)$ continues to indicate a large region of extended states around the band center. This behavior is qualitatively similar to what we observe in the classical-wave models and constitutes the first indication of a connection between ODD and localization of classical waves.

The diagonal-disorder (DD) case, panel d), follows the conventional Anderson scenario. Increasing disorder broadens the spectrum through fluctuations of the on-site energies and rapidly destroys the sharp structures present at weak disorder. Localization manifests itself through the progressive suppression of $\rho_{\mathrm{typ}}(\lambda)$ at both band edges, with localized regions expanding toward the band center as disorder increases. Unlike the ODD and classical-wave cases, no mechanism exists that protects an extended low-energy or central spectral sector. The resulting localization topology therefore differs qualitatively from that generated by the acoustic-sum-rule-constrained disorder classes.

Taken together, the DOS and TDOS results already reveal the central message of this work. While all four systems display disorder-induced localization, the way localization develops across the spectrum depends crucially on whether disorder acts within an unconstrained electronic Hamiltonian or within a constrained classical-wave operator obeying the acoustic sum rule. The full consequences of this distinction become apparent in the localization phase diagrams discussed next.

\subsection{Global phase diagrams and the topology of localized sectors}
The phase diagrams shown in Fig.~\ref{fig2} provide a unified representation of localization in the four disorder classes considered in this work. For each model, three characteristic lines are reported: the band edges (BE), obtained from the support of the arithmetic density of states (red); the mobility edges (ME) extracted from the support of the typical density of states $\rho_{\mathrm{typ}}(\lambda)$ (blue); and the corresponding estimate obtained from the participation ratio $\mathcal{P}$ (green). Overall, the mobility edges inferred from the typical DOS and the participation ratio show good qualitative agreement. Nevertheless, significant discrepancies emerge in certain regimes, most notably for the MD and DD models at strong disorder, and these will be examined in detail below.

A first observation is that increasing disorder enlarges the localized portions of the spectrum in all four models. The crucial differences concern the topology of this evolution: whether localization develops symmetrically or asymmetrically, whether extended states survive near the spectral center, and whether instability accompanies localization. The most striking distinction is that complete localization occurs only in the electronic DD model
\cite{bulka1987localization,grussbach95,dobro03},
corresponding to Anderson's original problem
\cite{anderson58}.
In contrast, the electronic ODD model and both classical-wave models retain a finite extended sector around the center of the spectrum even at the largest disorder strengths considered here. In these systems, localization remains confined to spectral regions near the band edges and complete localization is never reached~\cite{antoniou77,pinski12}.
\paragraph{Force-constant disorder.}
Panel~a) displays the phase diagram for force-constant disorder. At weak disorder the spectrum is bounded from below by $\omega^2=0$, and extended states occupy the central part of the stable spectrum. As disorder increases, localization first develops at large positive $\omega^2$ for $\Delta k\lesssim 3$, generating a stable localized branch near the upper band edge while the low-frequency sector remains extended. This asymmetry reflects the acoustic sum rule, which protects long-wavelength vibrational modes and suppresses localization in the low-frequency region.

For stronger disorder, $\Delta k\gtrsim3$, the spectrum extends into the region $\omega^2<0$, signaling the onset of mechanical instability. Importantly, instability and delocalization are distinct phenomena. The first unstable modes appear as a localized branch at negative $\omega^2$, while the central spectral region remains extended. Only at still larger disorder ($\Delta k\approx5$), where the unstable sector broadens substantially, unstable extended modes appear,
i.e. the mobility edge enters into the unstable regime.
The resulting phase diagram therefore contains three coexisting spectral sectors over a broad disorder range: an unstable localized branch at negative $\omega^2$, an extended central region, and a stable localized branch at large positive $\omega^2$.

The separation between instability and localization is a particularly interesting feature of the classical-wave problem. A related phenomenon has been observed in studies of the instantaneous-normal-mode spectrum of liquids. There, unstable modes are found to be fully localized at low temperatures, while a finite fraction becomes delocalized above a crossover temperature~\cite{berthier19}. Within the unstable-elastic description of these modes~\cite{schirm22,mossa23}, the disorder parameter plays a role analogous to temperature.
\paragraph{Mass disorder.}
The mass-disorder phase diagram, panel~b), shares with FCD the acoustic character of the spectrum but differs markedly in the extent of the localized high-frequency sector. We recall that positivity of the masses imposes the constraint $\Delta m\le2$, which defines the strongest physically admissible disorder. Unlike FCD, the spectrum remains strictly positive, $\omega^2\ge0$, and no unstable branch develops. The phase diagram therefore contains only stable localized and extended sectors. As in the FCD case, localization develops predominantly from the high-frequency side, while the low-frequency modes remain protected by the acoustic constraint.

\begin{figure}[t]
\centering
\includegraphics[width=0.5\textwidth]{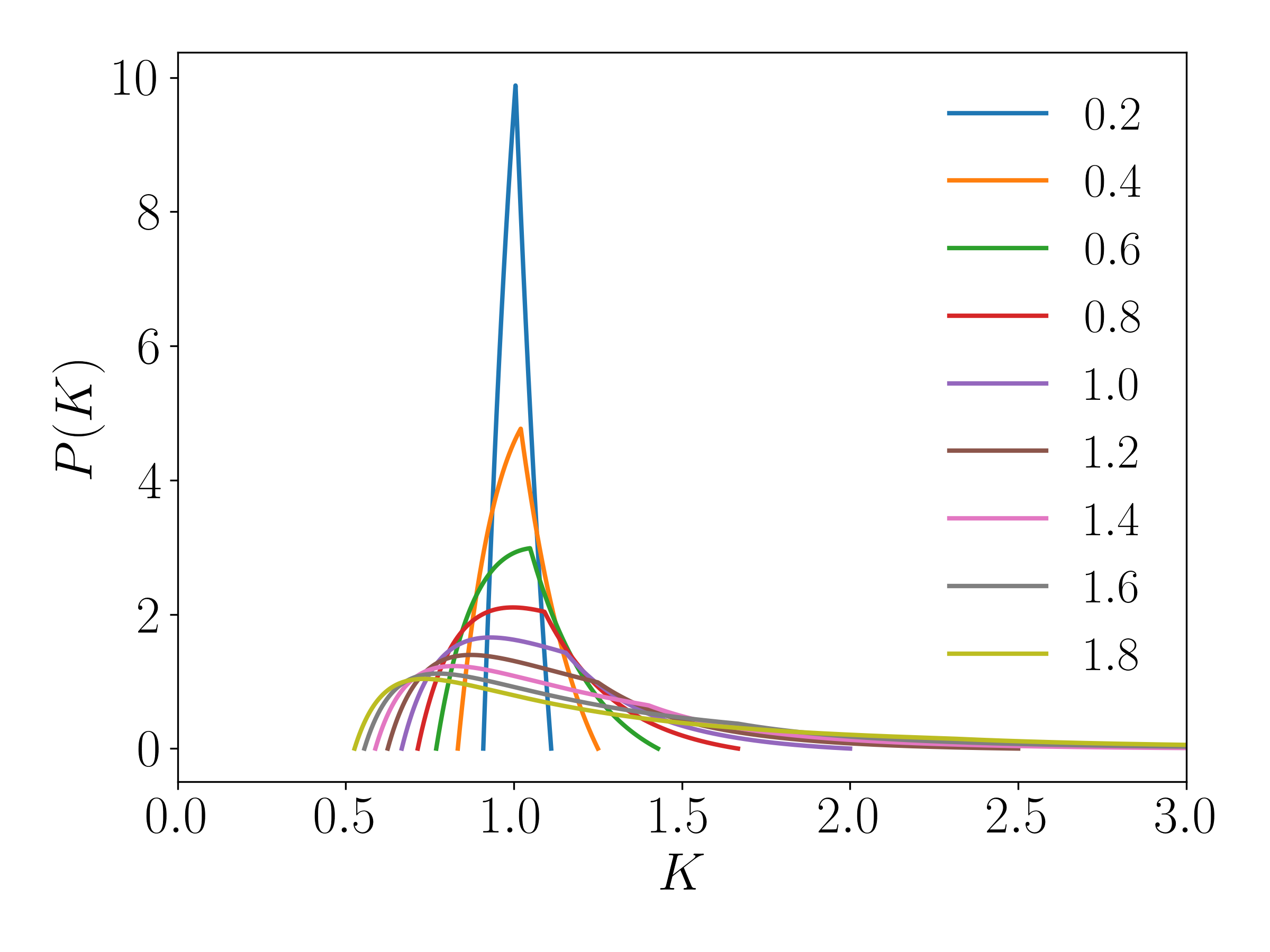}
\caption{Force-constant distribution $P(K)$ for mass disorder according to Eq.~(\ref{massdist2}). The labels indicate the values of the disorder parameters $\Delta m$.}
\label{massdist}
\end{figure}

This behavior can be understood by transforming the box distribution of masses, Eq.~(\ref{eq:mass-distr}), into an effective force-constant distribution. Setting $m_0=1$ and defining $K_{ij}=[m_i m_j]^{-1/2}\equiv K$, one obtains
\ba\label{massdist2}
P(K)&=&\frac{1}{\Delta m^2}\frac{2}{K^3}
\bigg[
\ln\left(\frac{2+\Delta m}{2-\Delta m}\right)
\nonumber\\
&&\qquad
-\bigg|
\ln\Bigg(
K^2
\bigg[1-\frac{1}{4}(\Delta m)^2\bigg]
\Bigg)
\bigg|
\bigg]
\nonumber\\
&&\times
\theta\!\left[
\left(K-\frac{1}{m_{\rm max}}\right)
\left(\frac{1}{m_{\rm min}}-K\right)
\right],
\ea
with $m_{\rm max}=1+\frac{1}{2}\Delta m$ and $m_{\rm min}=1-\frac{1}{2}\Delta m$. The corresponding distribution is bounded by $K_{\rm min}=m_{\rm max}^{-1}$ and $K_{\rm max}=m_{\rm min}^{-1}$. As shown in Fig.~\ref{massdist}, the box distribution of masses is transformed into a highly asymmetric effective force-constant distribution possessing a pronounced high-$K$ tail. As $\Delta m$ approaches its maximal value, this tail extends to arbitrarily large effective force constants. The resulting abundance of strongly coupled local environments naturally explains the broad localized high-frequency sector observed in the phase diagram. This feature constitutes one of the main results of the present work. The broad localized high-frequency regime is absent from the PT phase diagram reported in Ref.~\cite{pinski12} (cyan full circles in Fig.~\ref{fig2}). The re-entrant PT mobility edge closely follows the topology of the electronic DD phase diagram and therefore appears to be an artifact of the PT construction. We will go back to this point later.

A second noteworthy feature concerns the disagreement between the mobility edges extracted from the participation ratio and from the typical DOS at low frequencies. The participation-ratio analysis indicates the complete absence of localized low-frequency states throughout the accessible disorder range. By contrast, the typical DOS indicates a mobility edge that progressively invades the low-frequency region and ultimately spans the entire spectrum at $\Delta m=2$.

We regard the participation-ratio result as the more reliable one. Long-wavelength acoustic modes are only weakly scattered by disorder and therefore remain spatially extended over large length scales. In this regime finite-size effects become particularly severe, rendering the TDOS criterion delicate. The participation ratio, which is obtained directly from the eigenvectors, is less susceptible to this problem. We therefore conclude that the MD model contains localized states only in the high-frequency sector. This conclusion will be revisited and independently tested in the next subsection using level-statistics diagnostics.
\begin{figure*}
\centering
\includegraphics[width=1.0\textwidth]{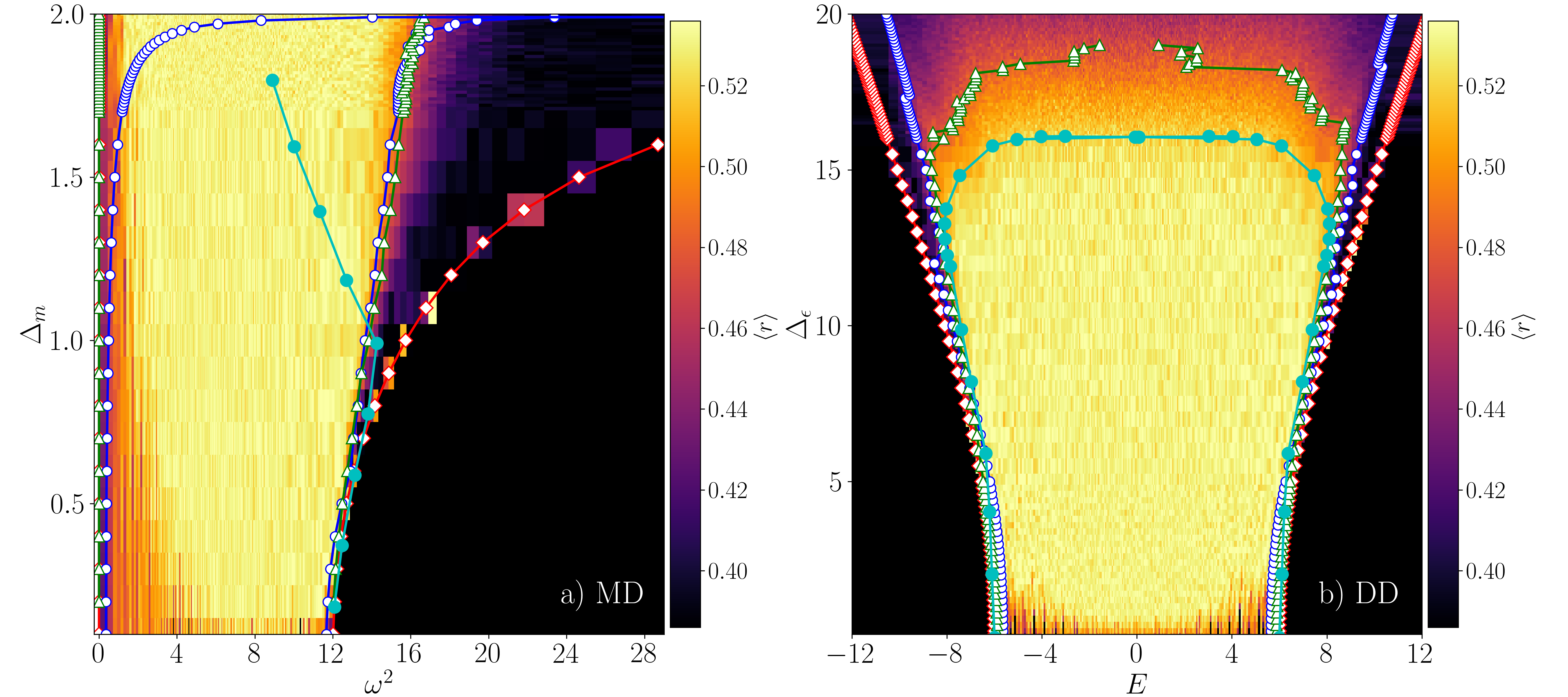}
\caption{Phase diagrams for classical mass disorder (MD) (panel a) ) and  electronic diagonal disorder (DD) (panel b), including the level-distance statistics. The color code signifies the value of $\langle r \rangle$, see legend. Black ($\langle r \rangle$ = 0.386) corresponds to complete localization with Poisson statistics, light yellow ($\langle r \rangle$ = 0.536) corresponds to complete delocalization. Red lines with white symbols are the band edges, Green lines with white symbols denote the mobility edges obtained from the participation ratio, blue lines with white symbols denote the mobility edges from the typical DOS. The cyan closed circles in the left panel are the mobility edges obtained with the PT approach using the transfer-matrix method~\cite{pinski12}. The cyan closed circles in the right panel are the mobility edges obtained with the transfer-matrix method in~\cite{bulka1987localization}.}
\label{fig4}
\end{figure*}
\paragraph{Off-diagonal disorder.}
Panel~c) shows the phase diagram for electronic off-diagonal disorder. Among the electronic models, this case is structurally closest to the classical-wave problem because disorder primarily enters the off-diagonal matrix elements. The phase diagram is symmetric about $E=0$, and localized sectors emerge symmetrically at the two spectral edges.

A notable feature is the weak non-monotonic behavior of the mobility edges at small disorder. Starting from the ordered limit, the mobility edges move slightly inward before reversing direction and steadily moving outward as $\Delta V$ increases
\cite{economou77,antoniou77,weaire77}.
Equivalently, after a very small initial reduction, the extended spectral region stabilizes while the localized sectors broaden from the spectral flanks.

The essential topological feature of ODD is therefore the persistence of a robust extended central sector. In contrast to DD, the mobility edges never approach the band center. In contrast to FCD, the two localized branches remain exactly equivalent because there is neither an acoustic sum rule nor an unstable sector. The phase diagram thus consists of a symmetry-protected extended region around $E=0$ flanked by two localized branches that expand symmetrically with increasing disorder.
\paragraph{Diagonal disorder.}
Panel~d) displays the conventional three-dimensional Anderson phase diagram. As expected \cite{bulka1987localization,grussbach95,dobro03}, the mobility edges 
start moving outwards with increasing disorder, then
move inward from both band edges as $\Delta\epsilon$ increases, progressively reducing the extent of the extended phase.

The crucial issue concerns the behavior near the band center. The mobility edge extracted from the participation ratio follows the expected Anderson scenario: the two branches continue moving inward and nearly close at $E=0$, indicating complete localization above a critical disorder strength. From Fig.~\ref{fig2}, this occurs at approximately $\Delta\epsilon_c\approx18\pm1$. By contrast, the mobility edge inferred from $\rho_{\mathrm{typ}}(\lambda)$ bends outward at large disorder, leaving a finite extended window around the band center.

We again regard the participation-ratio criterion as providing the physically more plausible result. Near the critical disorder, $\rho_{\mathrm{typ}}(\lambda)$ becomes extremely small and is obtained through a geometric average of the local DOS. As a consequence, it becomes particularly sensitive to finite-size effects, binning procedures, and statistical fluctuations. These difficulties are most severe near the critical region around $E=0$, where localization lengths are large and the distinction between an exponentially small and a truly vanishing TDOS becomes numerically subtle. The participation ratio is not affected by this particular source of instability and therefore provides a more reliable estimate of the disappearance of the extended phase.

The DD phase diagram should therefore be interpreted as the conventional Anderson scenario: localization develops symmetrically from both band edges, the extended region shrinks continuously with increasing disorder, and the mobility-edge branches eventually meet near the band center. In this respect DD differs fundamentally from ODD and from both classical-wave models. Although all three cases exhibit localization near the spectral boundaries, only DD undergoes complete localization, whereas the ODD, MD, and FCD models retain an extended central spectral sector throughout the accessible disorder range.
\subsection{Clarification of the MD and DD phase diagrams using level statistics}
Because the level-statistics analysis provides a third and fully independent probe of localization, it can be used to resolve the discrepancies between the mobility edges inferred from the participation ratio and from the typical DOS in the MD and DD models. To this end, Fig.~\ref{fig4} supplements the corresponding phase diagrams with color-coded values of the adjacent-gap ratio, $\langle r\rangle$, of Eq.~(\ref{eq:average_r}).

Fig.~\ref{fig4}~a) shows the phase diagram of the mass-disorder model together with the level-statistics data. The extended regime, characterized by values of $\langle r\rangle$ close to the GOE limit and highlighted by the yellow region, clearly extends down to $\omega^2=0$. We therefore conclude that no localized states exist in the low-frequency sector. This result is physically expected. As discussed in connection with Eq.~(\ref{massdist2}), the MD model can be interpreted as a particular form of force-constant disorder in which the effective force constants remain strictly positive. The low-frequency regime therefore corresponds to weakly scattered acoustic excitations undergoing Rayleigh scattering, whose mean free path diverges as $\omega^2\rightarrow0$.

It is instructive to compare the phase diagram obtained within the present MT framework with that reported by Pinski {\it et al.}~\cite{pinski12} (cyan closed circles), who employed the PT procedure. At weak disorder, the PT and MT mobility edges remain relatively close and follow the evolution of the upper band edge. However, for $\Delta m\gtrsim0.75$ the two descriptions diverge dramatically. The boundary of the localized regime obtained within the MT formulation, and independently confirmed by all three diagnostics employed here, differs qualitatively from the PT mobility edge. At the same time, the spectral support of the arithmetic DOS broadens strongly and, in fact, extends toward arbitrarily large frequencies as $\Delta m\rightarrow2$. As a consequence, an extended high-frequency band of localized states develops.

These observations constitute one of the principal results of the present work. First, the low-frequency acoustic sector remains extended throughout the physically accessible disorder range. Second, the correct MT formulation predicts a broad localized high-frequency regime that is entirely absent from the PT phase diagram.

Fig.~\ref{fig4}~b) shows the corresponding analysis for the electronic DD model. Here the level-statistics data clearly support the phase diagram obtained from the participation ratio. The resulting topology is precisely that envisioned in Anderson's original work~\cite{anderson58}: increasing disorder eventually destroys the central extended phase, leading to complete localization of the electronic spectrum.

The discrepancy between the participation-ratio and TDOS mobility edges can therefore be attributed to limitations of the typical-DOS criterion close to criticality. Near the localization transition, $\rho_{\mathrm{typ}}(\lambda)$ becomes extremely small and is obtained through a geometric average over strongly fluctuating local quantities. As a consequence, finite-size effects and statistical fluctuations become particularly severe in the critical region. By contrast, the participation ratio is computed directly from the eigenvectors and remains considerably more robust.

For comparison, Fig.~\ref{fig4}~b) also includes the mobility edge obtained for the same model from transfer-matrix calculations~\cite{bulka1987localization} (cyan solid circles). The overall topology agrees closely with the participation-ratio results and differs only slightly in the precise value of the critical disorder at $E=0$. It is worth noting that the typical DOS computed within the typical-medium approximation, an extension of the coherent-potential approximation~\cite{dobro03,dobro10}, also yields a conventional Anderson phase diagram with a critical disorder strength $\Delta\epsilon\approx16.3$, in good agreement with the transfer-matrix estimate. As the coherent-potential and
typical-medium approximations work in the thermodynamic limit, this indicates
that in this limit the typical DOS should exhibit the usual Anderson
scenario.

The combined evidence from Figs.~\ref{fig2} and~\ref{fig4} leads to a coherent picture of localization in the four disorder classes:

\begin{itemize}
\item
DD realizes the conventional Anderson scenario: two symmetric localized sectors emerge from the band edges and progressively eliminate the central extended phase until complete localization is reached.

\item
ODD also localizes from the spectral boundaries, but retains a robust extended region around $E=0$ throughout the accessible disorder range.

\item
MD exhibits a strongly asymmetric acoustic localization pattern. The low-frequency sector remains extended, whereas localization develops exclusively at high frequencies. As $\Delta m$ approaches its maximal value of 2, the upper boundary of the localized sector extends toward arbitrarily large frequencies. The resulting phase diagram differs qualitatively from that obtained within the PT framework.

\item
FCD combines acoustic asymmetry with mechanical instability and therefore exhibits the richest topology: a stable localized branch at large positive $\omega^2$, an extended central region, and an unstable localized branch at negative $\omega^2$.
\end{itemize}

The central lesson emerging from these phase diagrams is that the existence or absence of an extended spectral center provides the clearest distinction between the different disorder classes. More fundamentally, the results demonstrate that localization topology is controlled not only by disorder strength but also by the structural constraints built into the underlying eigenvalue problem. The acoustic sum rule therefore plays a role as important as disorder itself in determining the organization of localized and extended states within the spectrum.
\section{Discussion}
\label{sec:discussion}
We have presented a direct comparison of four localization scenarios: electronic tight-binding systems with diagonal disorder (DD) and off-diagonal disorder (ODD), and discretized scalar classical waves with mass disorder (MD) and force-constant disorder (FCD). The comparison was motivated by a long-standing question in the theory of wave localization: What is the correct electronic analogue of Anderson localization in disordered classical-wave systems? Our results show that the answer is more subtle than generally assumed.

The first central result concerns the distinction between the potential-type (PT) and modulus-type (MT) formulations of the classical-wave problem. The PT approach rewrites the wave equation in terms of an effective disorder potential and has often been used to establish an analogy with the electronic Anderson model with diagonal disorder. However, the resulting potential depends explicitly on the eigenvalue and therefore does not preserve the structure of the original spectral problem. By contrast, the MT formulation retains the correct linear eigenvalue structure and provides the appropriate framework for localization studies of classical waves.

The distinction is not merely formal. Applying the MT formulation to the mass-disorder problem yields a localization phase diagram that differs qualitatively from that obtained within the PT framework. In particular, we find a broad high-frequency localized regime whose extent grows dramatically as the disorder approaches the physical limit $\Delta m\rightarrow2$. This regime is absent from the PT phase diagram. The resulting MT phase diagram therefore establishes a qualitatively different localization topology for three-dimensional mass disorder.

A second central conclusion concerns the role of the acoustic sum rule. Once the classical-wave problem is formulated correctly within the MT framework, it becomes evident that disorder enters the operator in a fundamentally different way than in standard electronic Anderson models. The acoustic sum rule correlates diagonal and off-diagonal matrix elements and prevents them from being specified independently. Consequently, classical-wave systems cannot be identified with either diagonal or off-diagonal electronic disorder. Instead, they define a distinct constrained disorder class whose localization properties are controlled jointly by disorder strength and by the conservation laws embedded in the underlying wave operator.

This distinction is reflected directly in the topology of the localization phase diagrams. While the conventional DD model undergoes complete localization at sufficiently strong disorder, the ODD, MD, and FCD models retain an extended central spectral sector throughout the accessible disorder range. The classical-wave models additionally display a pronounced acoustic asymmetry: localization develops predominantly at high frequencies, whereas long-wavelength modes remain extended. In the FCD case this asymmetry coexists with mechanical instability, producing a phase-diagram topology with no counterpart in the standard electronic models.

An important aspect of the MD phase diagram is that it can be understood naturally in terms of the effective force-constant distribution generated by the mass-weighting transformation. As $\Delta m$ approaches its maximal value, the corresponding distribution develops a pronounced high-$K$ tail extending to arbitrarily large values. This generates a broad high-frequency sector of localized states while leaving the low-frequency acoustic modes largely unaffected.

A further outcome of the present work concerns the comparison of localization diagnostics. While the typical density of states, participation ratio, and level statistics generally provide consistent localization boundaries, important discrepancies arise in the MD and DD models. In both cases, the level-statistics analysis supports the mobility edges obtained from the participation ratio rather than those inferred from the typical DOS. These observations suggest that the participation ratio provides a particularly robust finite-size diagnostic in regimes where the typical DOS becomes extremely small and therefore especially sensitive to statistical fluctuations and finite-size effects.

The implications of these findings extend beyond the schematic scalar models studied here. The modulus-type formulation applies equally to electromagnetic waves, where fluctuations of dielectric permittivity and magnetic permeability play the role of modulus disorder. Our results therefore suggest that localization of three-dimensional classical waves should be sought in systems exhibiting strong spatial contrast in the relevant modulus-like quantities rather than in systems that can be interpreted as weak effective-potential perturbations. This conclusion is consistent with the recent study of disordered electromagnetic media with simultaneous fluctuations of permittivity $\varepsilon(\rr)$ and permeability $\mu(\rr)$~\cite{schirm24}, where an extended regime of localized states above the band edge of the disorder-free system was found, in close analogy with the high-frequency localized sector observed here for the MD model.

More generally, the present work suggests a different perspective on Anderson localization of classical waves. Rather than viewing acoustic, vibrational, and electromagnetic systems as imperfect realizations of the electronic Anderson model, they should be regarded as members of a broader family of localization problems in which conservation laws and structural constraints play a fundamental role. Within this framework, the acoustic sum rule emerges as a key organizing principle governing the distribution of localized and extended states throughout the spectrum.

We conclude that the most promising route toward observing Anderson localization of three-dimensional classical waves is likely to involve systems with strong modulus-type disorder, such as media with large mass-density contrasts in the acoustic case or correlated spatial fluctuations of permittivity and permeability in the electromagnetic case. More fundamentally, our results show that the topology of localization phase diagrams is controlled not only by disorder strength and dimensionality but also by the structural constraints embedded in the underlying wave operator. 

Three main conclusions therefore emerge. First, localization of classical waves must be analyzed within the modulus-type formulation rather than through the conventional potential-type mapping. Second, once formulated correctly, classical waves are seen to constitute a distinct constrained disorder class generated by the acoustic sum rule. Third, among the localization diagnostics considered here, the participation ratio, corroborated by level statistics, provides the most reliable determination of mobility edges in the challenging regimes of strong disorder and low-frequency acoustic excitations.
\begin{acknowledgments}
S. M. was supported by the project HeatFlow (Grant No. ANR-18-CE30-0019) funded by the French national funding agency, Agence Nationale de la Recherche.
\end{acknowledgments}
\section*{Data Availability}
The data that support the findings of this study are available from the corresponding author upon reasonable request.
\section*{Author Contributions}
S. M., G. R., and W. S. contributed equally to this work. 
\begin{figure*}[t]
\centering
\includegraphics[width=1\textwidth]{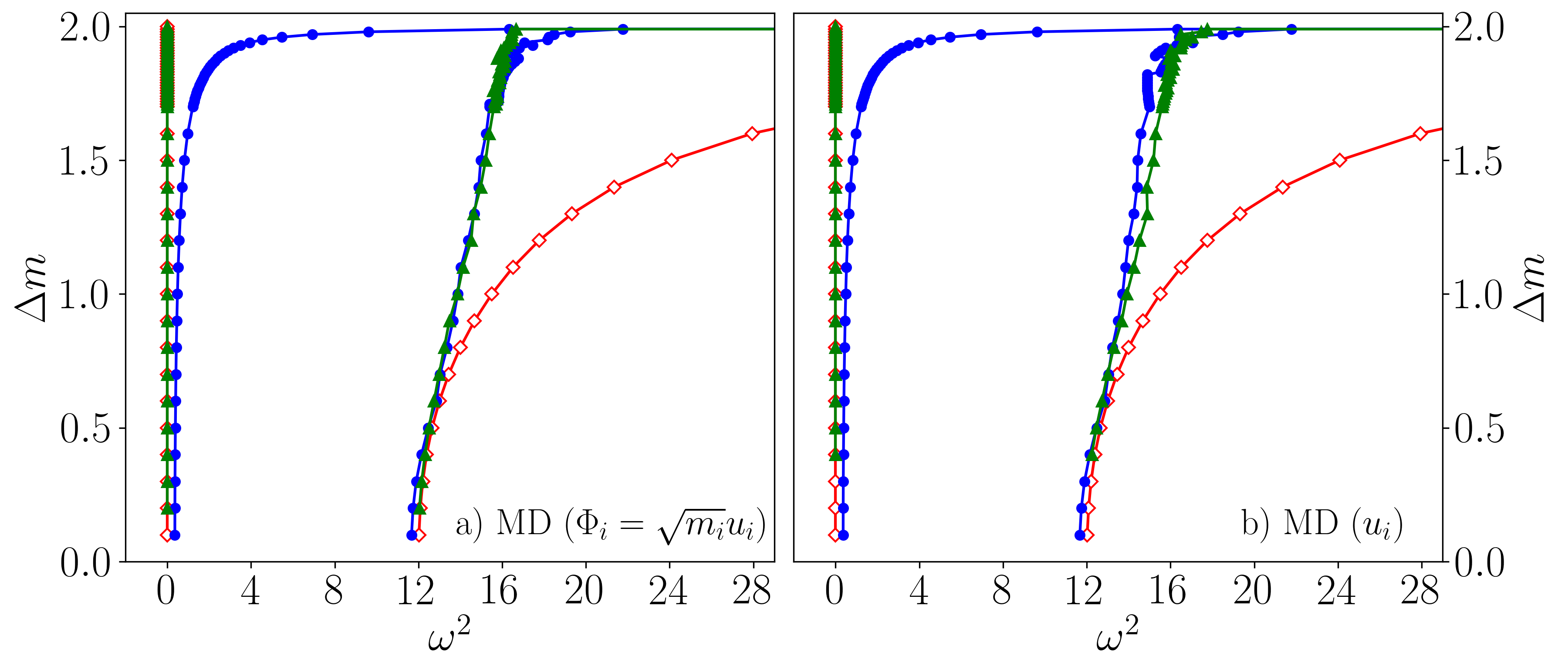}
\caption{Mass-disorder phase diagram obtained with two different eigenvector conventions. In panel (a) the observables are computed from the mass-weighted eigenvectors $\Phi_i=\sqrt{m_i}u_i$ that diagonalize the symmetric matrix; in panel (b) they are computed from the physical displacements $u_i$. The symbols are the same as in Fig.~\ref{fig2}.}
\label{ref:fig5}
\end{figure*}
\appendix
\setcounter{equation}{0}
\renewcommand\theequation{\thesection.\arabic{equation}}
\section{Two different eigenvector conventions for mass disorder}
In the mass-disorder case, the transformation $\phi_i=\sqrt{m_i}u_i$ was introduced to cast the eigenvalue problem into Hermitian form. As a consequence, two different fields are available for constructing localization diagnostics: the physical displacement field $u_i$ and the mass-weighted eigenvector $\phi_i$ used internally in the diagonalisation. To assess the influence of this choice, we computed the participation-ratio phase diagram using both representations. 

The results are shown in Fig.~\ref{ref:fig5}. While the low-frequency mobility edge remains essentially unchanged, the high-frequency branch exhibits a noticeable dependence on the chosen field: using $u_i$ yields a flatter family of mobility-edge lines at large $\Delta m$, whereas the participation-ratio boundary obtained from $\phi_i$ shifts more strongly with increasing disorder. Since localization is ultimately a property of the physical displacement field, this comparison indicates that participation-ratio diagnostics for mass disorder should be evaluated using $u_i$ rather than the auxiliary mass-weighted eigenvectors employed to restore Hermiticity.
\section{Discretization of the classical wave equations}
\paragraph{Force-constant disorder.}
Here, we derive the discretization for the one-dimensional cases. The generalization to three dimensions is straightforward. The classical wave equation for force-constant disorder is
\ba\label{a1}
-\omega^2u(x)&=&
\partial_x\bigg( K(x) \partial_x u(x)\bigg)\\
&=&\partial_x u(x)\partial_x K(x)+K(x)\partial_x^2u(x).\nonumber
\ea
We now discretize $\partial_xK(x)$ as
\be
\partial_xK(x)\rightarrow
\frac{1}{2h}
[K(x+h)
-K(x-h)]
\ee
and $\partial_x u(x)$ as
\be
\partial_x u(x)\rightarrow\frac{1}{a}[u(x+a)-u(x)]
\ee
or, equivalently,
\be
\partial_x u(x)\rightarrow\frac{1}{a}[u(x)-u(x-a)].
\ee
Here $a$ is the lattice constant, and $h$ an adjustable length of the order of $a$. We can now write
\ba
\partial_x u(x)\partial_x K(x)
&\rightarrow&\frac{1}{2h}[K(x+h)
-K(x-h)]\partial_x u(x)\nonumber\\
&\rightarrow&\frac{1}{2ha}\bigg(
K(x+h)[u(x+a)-u(x)]\nonumber\\
&&\!\!\!+K(x-h)[u(x-a)-u(x)\bigg),
\ea
while $\partial_x^2u(x)$ is discretized as
\be\label{e6}
\partial_x^2u(x)\rightarrow
\frac{1}{a^2}[u(x+a)+u(x-a)-2u(x)].
\ee
We can now set $h=\frac{1}{2}a$, obtaining
\begin{align}
\partial_x\bigg(K(x)\partial_x u(x)\bigg)
&\rightarrow
\frac{1}{a^2}\bigg(
[K(x+\tfrac{a}{2})+K(x)][u(x+a)-u(x)]
\nonumber\\
&\qquad
+[K(x-\tfrac{a}{2})+K(x)][u(x-a)-u(x)]\!\!\bigg).
\end{align}
If we define 
\be\label{e8}
u(x)\doteq u_i\qquad u(x\pm a)\doteq u_{i\pm 1},
\ee
and
\be
\frac{1}{a^2}\bigg[K(x\pm \frac{a}{2})+K(x)\bigg]\doteq K_{i,i\pm 1},
\ee
we obtain
\be
\partial_x\bigg( K(x) \partial_x u(x)\bigg)
\rightarrow
\sum_{j=i\pm 1}K_{ij}[u_j-u_i]
\ee
and, substituting in~(\ref{a1}), we obtain  Eq.~(\ref{eq:fcd_dyn})
\paragraph{Mass disorder.}
For the mass disorder case we have
\be
-\omega^2u(x)=\frac{1}{m(x)}\partial_x^2u(x).
\ee
Similarly to above, simply using (\ref{e6}), (\ref{e8}), and defining $m(x)\doteq m_i$, we obtain
\be
\frac{1}{m(x)}\partial_x^2u(x)
\rightarrow
\frac{1}{m_i}
\sum_{j=i\pm 1}[u_j-u_i],
\ee
recovering Eq.~(\ref{e14}).
\section{Discretization of the Schr\"odinger equation}
\paragraph{Potential disorder.}
The Schr\"odinger equation for potential disorder is ($\hbar =1$)
\be
E\psi(x,E)=\bigg(-\frac{1}{2m}\partial_x^2+\VV(x)\bigg)\psi(x,E)
\ee
Using Eq.~(\ref{e6}) and defining $\psi_i(E)\doteq\psi(x_i,E)$, we obtain the Anderson tight-binding equation
\be
E\psi_i(E)=\epsilon_i+\sum_{j=i\pm 1}V_0\psi_j(E),
\ee
with 
\be
\epsilon_i=\frac{1}{a^2}\VV(x_i),
\ee
and
\be
V_0=-\frac{1}{2ma^2}\,.
\ee
\paragraph{Effective-mass disorder.}
As shown in~\cite{levyleblond95}, the Schr\"odinger equation for an electron with spatially varying effective mass reads
\be
E\psi(x,E)=\bigg(-\frac{1}{2}\partial_x\frac{1}{m(x)}\partial_x+\VV(x)\bigg)\psi(x,E).
\ee
We recognize that the effective-mass term has the same form as the modulus term in~(\ref{a1}). We therefore immediately obtain, 
{\it mutatis mutandis}, the discretized version
\be
E\psi_i(E)=\epsilon_i+
\sum_{j=i\pm 1}V_{ij}
\psi_j(E),
\ee
with
\be
\epsilon_i=\frac{1}{a^2}\VV(x_i)-
\sum_{j=i\pm 1}
V_{ij},
\ee
and
\be
V_{i,i\pm 1}=
-\frac{1}{2a^2}\bigg[\frac{1}{m(x\pm \frac{a}{2})}
-\frac{1}{m(x)}\bigg]\, .
\ee
\bibliography{typical}
\end{document}